\documentclass[english]{article}
\usepackage[T1]{fontenc}
\usepackage[latin9]{inputenc}
\usepackage{geometry}
\usepackage{xcolor}
\usepackage{pdfcolmk}
\usepackage{amsmath}
\usepackage{amsthm}
\usepackage{amssymb}
\usepackage{graphicx}
\usepackage[numbers]{natbib}
\PassOptionsToPackage{normalem}{ulem}
\usepackage{ulem}

\makeatletter

\providecolor{lyxadded}{rgb}{0,0,1}
\providecolor{lyxdeleted}{rgb}{1,0,0}
\DeclareRobustCommand{\lyxsout}[1]{\ifx\\#1\else\sout{#1}\fi}

\numberwithin{equation}{section}
\numberwithin{figure}{section}
\newcommand{\lyxaddress}[1]{
	\par {\raggedright #1
	\vspace{1.4em}
	\noindent\par}
}

\usepackage{hyperref}

\makeatother

\usepackage{babel}
\begin{document}
\title{The full analytic trans-series in integrable field theories}
\author{Zoltán Bajnok$^{1}$, János Balog$^{1}$ and István Vona$^{1,2}$}
\maketitle

\lyxaddress{\begin{center}
\emph{1. Wigner Research Centre for Physics,}\\
\emph{Konkoly-Thege Miklós u. 29-33, 1121 Budapest , Hungary}\\
\emph{and}\\
\emph{2. Roland Eötvös University, }\\
\emph{ Pázmány Péter sétány 1/A, 1117 Budapest, Hungary}
\par\end{center}}
\begin{abstract}
We analyze a family of generalized energy densities in integrable
quantum field theories in the presence of an external field coupled
to a conserved charge. By using the Wiener-Hopf technique to solve
the linear thermodynamic Bethe ansatz equations we derive the full
analytic trans-series for these observables in terms of a perturbatively
defined basis. We show how to calculate these basis elements to high
orders analytically and reveal their complete resurgence structure.
We demonstrate that the physical value of the energy density is obtained
by the median resummation of the perturbative series.
\end{abstract}

\section{Introduction}

Systems soluble by the Bethe ansatz are relevant in condensed matter
systems, in statistical as well as in particle physics \citep{Samaj:2013yva,Mussardo:2020rxh}.
They provide toy models, where non-perturbative, strongly interacting
phenomena such as dynamical mass generation, superconductivity, etc.
can be analyzed in simplified circumstances. Additionally, some of
them also have experimental realizations.

In most of the applications we are interested in the thermodynamic
limit of the Bethe ansatz equations, which can be formulated as linear
integral equations. Unfortunately, these integral equations cannot
be solved exactly and we have to rely on their perturbative expansions.
These perturbative series are asymptotic and their factorial growth
signals non-perturbative terms. For a complete description one has
to build a multiple series, i.e. a trans-series both in the perturbative
coupling and in the exponentially suppressed non-perturbative corrections.
This trans-series is understood as Borel resummed, and the requirement
of being free of ambiguities requires an intricate interplay between
the various perturbative and non-perturbative terms\footnote{See \citep{Aniceto:2013fka} and references therein for the ambiguity
cancellation}. The theory which formulates this is called resurgence, which lives
its renaissance now, see \citep{Marino:2012zq,Dorigoni:2014hea,Aniceto:2018bis}
for recent reviews.

Recently there has been great progress in the perturbative as well
as \emph{the leading non-perturbative} analysis of the linear TBA
equations. On the condensed matter side the groundstate energy density
of the Lieb-Liniger, Gaudin-Yang and Hubbard models together with
their generalizations were investigated \citep{Marino:2019fuy,Marino:2019wra,Marino:2020dgc,Marino:2020ggm,PhysRevA.106.062216,PhysRevLett.130.020401}.
The non-pertubative terms were in many cases related to the superconductive
gap as well as to renormalon diagrams. On the particle physics side
free energies of asymptotically free integrable quantum field theories
in the presence of an external field coupled to a conserved charge
were analyzed \citep{Polyakov:1983tt,Hasenfratz:1990ab,Hasenfratz:1990zz,Forgacs:1991rs,Balog:1992cm,Evans:1994sv,Evans:1994sy}.
These included the $O(N)$ non-linear sigma model and its supersymmetric
extension, the Gross-Neveu model and the principal chiral field for
which the large order behavior of the perturbative series were also
investigated \citep{Marino:2021dzn,Bajnok:2021dri,Bajnok:2021zjm,Marino:2022ykm,Bajnok:2022rtu}.
In these quantum theories the non-perturbative terms are related to
instantons or renormalons, which were further confirmed by large $N$
calculations \citep{Marino:2021six,DiPietro:2021yxb} and in the $O(3)$
model by introducing a $\theta$-term \citep{Marino:2022ykm}.

In constructing the ambiguity free trans-series the first problem
is to efficiently calculate the perturbative terms. This was first
achieved for the energy density of $O(N)$ models in \citep{Volin:2009wr,Volin:2010cq}.
This method was extended for statistical models and for the circular
plate capacitor \citep{Marino:2019fuy,Reichert:2020ymc} and by combining
with the Wiener-Hopf technique to integrable quantum field theories
\citep{Marino:2019eym}. The first few exponentially suppressed corrections
can be extracted from the asymptotics of the perturbative coefficients
\citep{Marino:2019fuy,Marino:2019eym,Marino:2019wra,Marino:2020dgc,Abbott:2020mba,Abbott:2020qnl,Bajnok:2021dri,Bajnok:2021zjm,Bajnok:2022rtu}.
A systematic treatment based on the Wiener-Hopf approach was presented
in \citep{Marino:2021dzn}, which resulted in the precise structure
of the trans-series and explicit calculations of the first few non-perturbative
corrections. This was further extended to higher orders and improved
by introducing the running coupling for the $O(N)$ models in \citep{Bajnok:2022rtu}.
The aim of our present paper is to\emph{ solve completely} these models
by determining the full trans-series, i.e. all the non-perturbative
terms together with their perturbative expansions. We are doing this
by expressing these higher perturbative expansions in terms of the
original perturbative series of generalized observables, which we
also determine from the known perturbative series of the energy density.

The paper is organized as follows. In section 2 we introduce the integral
equation, the generalized observables and differential equations which
relate them to each other. In section 3 we demonstrate how the Wiener-Hopf
technique can be used to calculate these generalized observables.
This provides a structural result, which we make explicit by introducing
a perturbatively calculable basis in section 4 and constructing the
full trans-series. In section 5 we present a method for determining
the basis and investigate how the various parts of the trans-series
are connected. We also relate its median resummation to the TBA result.
We provide explicit examples in section 6. Finally, we conclude in
section 7.

\section{Observables and their properties}

We investigate linear integral equations of the form
\begin{equation}
\chi_{n}(\theta)-\int_{-B}^{B}d\theta'K(\theta-\theta')\chi_{n}(\theta')=r_{n}(\theta)\label{eq:TBA}
\end{equation}
for $\vert\theta\vert\leq B,$ where $r_{n}(\theta)=\cosh(n\theta)$
and the kernel is a symmetric function, which, in most of the applications,
is related to the logarithmic derivative of the scattering matrix
\citep{Polyakov:1983tt,Hasenfratz:1990ab,Hasenfratz:1990zz,Forgacs:1991rs,Balog:1992cm,Evans:1994sv,Evans:1994sy,Marino:2019fuy}.
We are interested in the observables 
\begin{equation}
{\cal O}_{n,m}=\int_{-B}^{B}\frac{d\theta}{2\pi}\chi_{n}(\theta)r_{m}(\theta)
\end{equation}
as functions of $B$, but we do not indicate this dependence explicitly.
This observable is symmetric in $n$ and $m$, which are not necessarily
integers but non-negative. Its $B$-derivative (which we denote by
a dot) can be written in terms of the boundary values of $\chi_{n}$-s
as \citep{Bajnok:2022rtu}
\begin{equation}
\frac{d{\cal O}_{n,m}}{dB}\equiv\dot{{\cal O}}_{n,m}=\frac{1}{\pi}\chi_{n}(B)\chi_{m}(B)\quad.\label{eq:Onmdot}
\end{equation}
By generalising the manipulation of the integral equation in \citep{PhysRevLett.130.020401,PhysRevA.106.062216}
one can show that these boundary values satisfy the differential equation
\begin{equation}
\frac{\ddot{\chi}_{n}(B)}{\chi_{n}(B)}-n^{2}=f(B)\quad,\label{eq:chidot}
\end{equation}
where $f(B)$ is an $n$-independent function, which can be calculated,
for instance, from the $n=1$ case. These equations connect all observables
to one of them, say to ${\cal O}_{1,1}$, which is the groundstate
energy of the integrable model in a magnetic field coupled to a conserved
charge. The observable ${\cal O}_{n,m}$ for $n,m$ integers can be
interpreted as the expectation value of the conserved spin $m$ charge,
in the presence of the magnetic field, when the Hamiltonian is given
by the conserved spin $n$ charge and we have periodicity in the direction
of the corresponding generalized volume. Similar formulae appear in
the generalized hydrodynamics for expectation values of conserved
charges and currents, although in slightly different settings \citep{Castro-Alvaredo:2016cdj,Bajnok:2019mpp}

Let us note that although we analyse the case of the $r_{n}(\theta)=\cosh(n\theta)$
source terms, they can be used to express the solutions and observables
for the $r_{n}(\theta)=\sinh(n\theta)$ source terms as well. They
together form a complete basis, from which by differentiation or analytical
continuation and Fourier transform cases with more general sources
can be reached \citep{Bajnok:2022prepi}.

\section{Wiener-Hopf integral equation}

The standard way to solve the integral equation is the Wiener-Hopf
technique \citep{Hasenfratz:1990ab,Hasenfratz:1990zz,Marino:2021dzn,Bajnok:2022rtu}.
As a first step we extend the source as $r_{n}(\theta)=\Theta(-\theta+B)\frac{e^{n\theta}}{2}+\Theta(\theta+B)\frac{e^{-n\theta}}{2}$
, ($\Theta$ being the Heavyside theta function), as well as the integrations,
(but not $\chi_{n}(\theta)$), for the whole line
\begin{equation}
\chi_{n}(\theta)-\int_{-\infty}^{\infty}d\theta'K(\theta-\theta')\chi_{n}(\theta')=r_{n}(\theta)+R(\theta)+R(-\theta)
\end{equation}
by paying the price of introducing an unknown function $R(\theta)$,
which, however, vanishes for $\theta<B$. In solving the equation
in Fourier space the key point is the factorization 
\begin{equation}
(1-\tilde{K}(\omega))^{-1}=G_{+}(\omega)G_{+}(-\omega)
\end{equation}
into factors analytic in the lower and upper half planes. This can
be done by taking logarithms and projecting into the appropiate analytic
pieces \citep{Bajnok:2022rtu}. Implementing the separation of the
equation into lower and upper half analytical pieces we arrive at
\begin{equation}
X_{n}(i\kappa)+\int_{-\infty}^{\infty}\frac{e^{2i\omega B}\sigma(\omega)X_{n}(\omega)}{\kappa-i\omega}\frac{d\omega}{2\pi}=\frac{1}{n-\kappa}\quad,
\end{equation}
where $\sigma(\omega)=\frac{G_{+}(-\omega)}{G_{+}(\omega)}$ and the
unknown function $X_{n}(\omega)$ is related to the Fourier transform
of $R(\theta)$ as 
\begin{equation}
X_{n}(\omega)=\frac{2e^{-(n+i\omega)B}G_{+}(\omega)\tilde{R}(\omega)}{G_{+}(in)}+\frac{G_{+}(\omega)}{G_{+}(in)}\frac{1}{(n+i\omega)}\quad.
\end{equation}
Except for the explicitly introduced pole at $\omega=in$, $X_{n}(\omega)$
is analytic in the upper half plane. We also assume that $n>0$. The
$n=0$ case requires special care \citep{Marino:2021dzn,Bajnok:2022rtu}
and we can recover it by solving the differential equations (\ref{eq:Onmdot},\ref{eq:chidot}).

In the typical applications $\sigma(\omega)$ has a cut and poles
at $\omega=i\kappa_{l},\,l=1,2,\dots,$ on the positive imaginary
line. Additionally, we also have the explicit pole at $\omega=in$
coming from $X_{n}$. It is advantageous to disentangle the poles
from the cut by moving the cut a bit away from the imaginary line
in either direction \citep{Marino:2021dzn,Bajnok:2022rtu}. We then
deform the integration contour from the real line surrounding the
cut and separately the poles whose residues we collect. We can do
it in two different ways, by integrating a bit left or right of the
poles. The residues will also depend on this choice, but the final
result must be the same. This is the manifestation of a Stokes phenomena
and the two different choicesare related to the two lateral resummations.
This is by no means obvious and we have only a confirmation a posteriori.
Indeed, by switching between the contours the residue terms providing
the Stokes constants change signs, which is the same which comes out
from the lateral Borel resummations of the trans-series solution of
the problem. For definiteness, we integrate a bit left of the imaginary
line ( in $\kappa$ a bit above the real positive line): 
\begin{align}
X_{n}(i\kappa)+i\sum_{l=0}^{\infty}\frac{S_{l}q_{n,\kappa_{l}}}{\kappa+\kappa_{l}}e^{-2\kappa_{l}B}\qquad\\
+\int_{C_{+}}e^{-2B\kappa'}\frac{\delta\sigma(i\kappa')X_{n}(i\kappa')}{\kappa+\kappa'}\frac{d\kappa'}{\pi} & =\frac{1}{n-\kappa}\quad,\nonumber 
\end{align}
where $q_{n,\kappa_{l}}=X_{n}(i\kappa_{l})$ and $S_{l}$ is the residue
of $i\sigma(i\kappa+0)$ at $\kappa=\kappa_{l}$, while $\delta\sigma(\kappa)=\frac{1}{2i}(\sigma(i\kappa-0)-\sigma(i\kappa+0))$
is the discontinuity of $\sigma$. We included in the sum the contribution
of the pole of $X_{n}(i\kappa)$ at $\kappa_{0}=n$ with residue $S_{0}=-i\sigma(in+0)=-i\sigma_{n}^{+}$
with the convention that $q_{n,n}=1$. Here we assume that all poles
$\kappa_{l}$ are distinct, including $\kappa_{0}$, i.e. $n\neq\kappa_{l}$.
In the more general case, $\sigma(i\kappa)$ can have higher order
poles, which could even coincide with the pole at $\omega=in$. In
this case $S_{l}$ and $q_{n,\kappa_{l}}$ are related to the expansion
of the functions around the singularity, however, we do not consider
these complicated cases in this short letter, see our upcoming paper
for further details \citep{Bajnok:2022prepi}.

Typically, we can introduce a running coupling $v$ 
\begin{equation}
\kappa=vx\quad;\qquad2B=v^{-1}+\gamma\log v+L\label{eq:runningcoupling}
\end{equation}
with an arbitrary constant $L$, such that the integral equation for
the rescaled variable $Q_{n}(x)=X_{n}(ivx)$ takes the generic form
\begin{align}
Q_{n}(x)+i\sum_{l=0}^{\infty}\frac{S_{l}q_{n,\kappa_{l}}\nu^{\kappa_{l}}}{\kappa_{l}+vx}\qquad\label{eq:Qeq}\\
+\int_{C_{+}}\frac{e^{-y}{\cal A}(y)Q_{n}(y)}{x+y}\frac{dy}{\pi} & =\frac{1}{n-vx}\quad,\nonumber 
\end{align}
where 
\begin{align}
q_{n,\kappa_{j}}=Q(\frac{\kappa_{j}}{v})= & -i\sum_{l=0}^{\infty}\frac{S_{l}q_{n,\kappa_{l}}\nu^{\kappa_{l}}}{\kappa_{l}+\kappa_{j}}\nonumber \\
 & -\int_{C_{+}}\frac{e^{-y}{\cal A}(y)Q_{n}(y)}{\kappa_{j}+vy}\frac{vdy}{\pi}+\frac{1}{n-\kappa_{j}}\quad,\label{eq:qeq}
\end{align}
and $q_{n,n}=1$ while 
\begin{equation}
\nu=e^{-2B}=e^{-L}v^{-\gamma}e^{-1/v}
\end{equation}
The model-dependent parameter $\gamma$ has to be chosen such that
${\cal A}(y)=e^{-vy(\gamma\log v+L)}\delta\sigma(vy)$ has a power-series
expansion in $v$ without any $\log v$ terms: ${\cal A}(y)=\sum_{j=0}^{\infty}v^{j}\alpha_{j}(y)$.
By appropriately choosing $L$ the linear $y$-dependence in $\log{\cal A}(y)$
can be canceled. Here $q_{n,\kappa_{l}}$-s (except $q_{n,n}=1$)
are also unknowns, which have to be calculated by evaluating the integral
equation \eqref{eq:Qeq} at the positions $xv=\kappa_{l}$. If $Q_{n}(x)$
including $q_{n,\kappa_{l}}$ are determined, then the observable
${\cal O}_{n,m}$ can be written (similarly to \citep{Bajnok:2022rtu})
as
\begin{equation}
{\cal O}_{nm}=\frac{e^{(n+m)B}}{4\pi}G_{+}(im)G_{+}(in)W_{n,m}\label{eq:Onm}
\end{equation}
with 
\begin{align}
W_{n,m} & =\frac{1}{n+m}+i\sum_{l=0}^{\infty}\frac{S_{l}q_{n,\kappa_{l}}\nu^{\kappa_{l}}}{m-\kappa_{l}}+\sigma_{m}^{+}\nu^{m}q_{n,m}\nonumber \\
 & \quad+\frac{v}{\pi}\int_{C_{+}}\frac{e^{-x}{\cal A}(x)Q_{n}(x)}{m-vx}dx\quad,
\end{align}
where $q_{n,m}=Q_{n}(\frac{m}{v})$. Here we assumed that $n\neq\kappa_{l}$
and $n\neq m$, otherwise we have to calculate the residue of a second
order pole. The $n=m$ case can be recovered by taking the $n\to m$
limit. For the boundary value of the field, similarly to \citep{Bajnok:2022rtu},
we obtain 
\begin{equation}
\chi_{n}(B)=\frac{e^{nB}}{2}G_{+}(in)w_{n}\label{eq:chin}
\end{equation}
with
\begin{equation}
w_{n}=1+i\sum_{l=0}^{\infty}S_{l}q_{n,\kappa_{l}}\nu^{\kappa_{l}}+\frac{v}{\pi}\int_{C_{+}}e^{-x}{\cal A}(x)Q_{n}(x)dx
\end{equation}
for $n\text{\ensuremath{\neq0}. }$ In the case of $n=0$ we need
to calculate $f$ from $\chi_{1}$ in \eqref{eq:chidot} and solve
the equation \eqref{eq:chidot} for $\chi_{0}$ and \eqref{eq:Onmdot}
for ${\cal O}_{0,0}$.

\section{Trans-series ansatz and its solution}

We solve the equations (\ref{eq:Qeq},\ref{eq:qeq}) for $Q_{n}(x)$
in terms of a trans-series ansatz
\begin{equation}
Q_{n}(x)=\sum_{l=0}^{\infty}\nu^{d_{l}}\sum_{j=0}^{\infty}Q_{n,j}^{(d_{l})}(x)v^{j}\quad,\label{eq:Qntransansatz}
\end{equation}
where the set of nonzero $Q_{n,j}^{(d_{l})}$-s is model-dependent
and encodes via $\nu\sim e^{-1/v}$ the non-perturbative corrections.
One has to investigate the set of poles $\{\kappa_{l}\}$, which (up
to some isolated cases such as $\kappa_{0}$) can be described as
a union of finitely many sets of the form $\{a_{i}l+b_{i}\}$ with
$l=1,2,\dots$. We should introduce $d_{l}$ such that all non-peturbative
corrections are accounted for. We give concrete examples in section
7.

The generic solution to \eqref{eq:Qntransansatz} can be calculated
iteratively in $l$. We start with the $l=0$ perturbative part, i.e.
we have to solve perturbatively the following problem:
\begin{equation}
P_{\alpha}(x)+\int_{C_{+}}\frac{e^{-y}{\cal A}(y)P_{\alpha}(y)}{x+y}\frac{dy}{\pi}=\frac{1}{\alpha-vx}\quad.
\end{equation}
This can be done by expanding ${\cal A}(y)$ and the source term in
power series in $v$ and iteratively solving at any order based on
lower order solutions \citep{Marino:2021dzn,Bajnok:2022rtu}. We will
see, however, that the explicit solution is not needed. Observe also
that originally we needed $P_{\alpha}$ for $\alpha>0$, but the equation
and the perturbative solution make perfect sense also for $\alpha<0$.
Using these solutions the unknown $Q_{n}(x)$ can be written as 
\begin{equation}
Q_{n}(x)=P_{n}(x)+i\sum_{l=0}^{\infty}S_{l}q_{n,\kappa_{l}}\nu^{\kappa_{l}}P_{-\kappa_{l}}(x)\quad,\label{eq:Qn}
\end{equation}
where, from the definition of $q_{n,\kappa_{s}}$, we obtain a closed
system of linear equations of the form 
\begin{equation}
q_{n,\kappa_{s}}-i\sum_{l=0}^{\infty}S_{l}q_{n,\kappa_{l}}\nu^{\kappa_{l}}A_{-\kappa_{l},-\kappa_{s}}=A_{n,-\kappa_{s}}\label{eq:qn}
\end{equation}
with the exception of $q_{n,n}=1$. Here we introduced the symmetric
building block (for $\alpha\neq-\beta$) as 
\begin{equation}
A_{\alpha,\beta}=\frac{1}{\alpha+\beta}+\langle P_{\alpha}\rangle_{\beta}\quad,
\end{equation}
which contains the moment 
\begin{equation}
\langle Q\rangle_{\beta}=\int_{C_{+}}\frac{e^{-x}{\cal A}(x)Q(x)}{\beta-vx}\frac{vdx}{\pi}\quad.
\end{equation}
The symmetric moments $\langle P_{\alpha}\rangle_{\beta}$ are understood
perturbatively in $v$ and are well-defined for any signs of $\alpha$
and $\beta$. We note that the recursive structure for $q_{n,\kappa_{l}}$
is the consequence of the integral equation, where the model-specific
feature lies in the set $\kappa_{l}$ (the non-perturbative nature)
as well as in $A_{n,m}$ (the perturbative nature). In the following
we solve this linear system of equations. Since $q_{n,n}=1$ is not
an unknown, we regard its contribution as an inhomogeneous source
term
\begin{equation}
q_{n,\kappa_{s}}-i\sum_{l=1}^{\infty}q_{n,\kappa_{l}}S_{l}\nu^{\kappa_{l}}A_{-\kappa_{l},-\kappa_{s}}=s_{n,-\kappa_{s}}
\end{equation}
where $s_{n,-\kappa_{s}}=A_{n,-\kappa_{s}}+\sigma_{n}^{+}\nu^{n}A_{-n,-\kappa_{s}}$.
This linear matrix equation $(\mathbb{\mathbf{I}-\mathbf{A}})\mathbf{q}_{n}=\mathbf{s}_{n}$,
with $\mathbf{A}_{s,l}=iS_{l}\nu^{\kappa_{l}}A_{-\kappa_{s},-\kappa_{l}}$
can be solved by inversion $\mathbf{q}_{n}=(\mathbb{\mathbf{I}-\mathbf{A}})^{-1}\mathbf{s}_{n}$,
which can be represented by the Neumann series $\mathbf{q}_{n}=(\mathbf{I}+\mathbf{A+\mathbf{A^{2}+\dots}})\mathbf{s}_{n}$
and expanded in $\nu$. Alternatively, we can plug back recursively
every lower order solution in to the $\nu$-expansion, leading to
\begin{equation}
q_{n,\kappa_{s}}=\sum_{\mathrm{paths}}sA_{n,\mathrm{path},-\kappa_{s}}S_{\mathrm{path}}\quad,\label{eq:qnmsol}
\end{equation}
where a path means a sequence starting from $n$ and ending at $-\kappa_{s}$:
$(n,l_{1},l_{2},\dots,l_{N},-\kappa_{s})$. The contribution of such
a path is 
\begin{align}
sA_{n,\mathrm{path},-\kappa_{s}} & =s_{n,-\kappa_{l_{1}}}A_{-\kappa_{l_{1}},-\kappa_{l_{2}}}\dots A_{-\kappa_{l_{N}},-\kappa_{s}}\nonumber \\
S_{\mathrm{path}} & =iS_{l_{1}}\nu^{\kappa_{l_{1}}}\dots iS_{l_{N}}\nu^{\kappa_{l_{N}}}\qquad,
\end{align}
where the inner indices take only the values $l_{k}=1,2,\dots$. At
each non-perturbative order in $\nu$ we have only a finite number
of terms contributing. With this solution the unknown function $Q_{n}\left(\frac{m}{v}\right)=q_{n,m}$
can also be written as 
\begin{equation}
q_{n,m}=A_{n,-m}+i\sum_{l=0}^{\infty}S_{l}q_{n,\kappa_{l}}\nu^{\kappa_{l}}A_{-\kappa_{l},-m}\quad.
\end{equation}

The observables $W_{n,m}$ can be obtained in terms of $q_{n,\kappa_{l}}$
as 
\begin{align}
W_{n,m} & =s_{m,n}+i\sum_{l=0}^{\infty}S_{l}q_{n,\kappa_{l}}s_{m,-\kappa_{l}}\nu^{\kappa_{l}}=A_{n,m}+O(\nu).\label{eq:Wnm}
\end{align}
Clearly, the basic building block $A_{n,m}$ is nothing but the perturbative
part of our generic observable $W_{n,m}$. The boundary value of the
field can be expressed as 
\begin{equation}
w_{n}=a_{n}+i\sum_{l=0}^{\infty}S_{l}q_{n,\kappa_{l}}\nu^{\kappa_{l}}a_{-\kappa_{l}}\quad,\label{eq:wn}
\end{equation}
where $a_{\alpha}=\lim_{\beta\to\infty}\beta A_{\alpha,\beta}$.

By this we provided a complete recursive solution of the problem,
i.e. we expressed the observables in terms of the perturbatively defined
$A_{n,m}$-s. In the following we explain how the perturbative expansion
of the building blocks can be calculated.

\section{Median resummation and alien derivatives}

Let us summarize what we have achieved so far. The observables ${\cal O}_{n,m}$
and $\chi_{n}(B)$ can be written in terms of $W_{n,m}$ and $w_{n}$
as \eqref{eq:Onm} and \eqref{eq:chin}, which satisfy two differential
equations 
\begin{align}
(n+m)W_{n,m}+\dot{W}_{n,m} & =w_{n}w_{m}\label{eq:de1}\\
2n\dot{w}_{n}+\ddot{w}_{n} & =fw_{n}\label{eq:de2}
\end{align}
and have the solutions \eqref{eq:Wnm} and \eqref{eq:wn} in terms
of $q_{n,\kappa_{s}}$, which is given by \eqref{eq:qnmsol}.

The perturbative parts $A_{n,m}$ and $a_{n}$ satisfy the $W\to A,w\to a$
differential equations (\ref{eq:de1},\ref{eq:de2}). Since Volin's
method \citep{Volin:2009wr,Volin:2010cq,Marino:2019eym} determines
$W_{1,1}$ at the perturbative level it provides $A_{1,1}$. We can
then extract the perturbative part of $w_{1}$, namely $a_{1}$ from
\eqref{eq:de1}, and by plugging back to eq. \eqref{eq:de2} we can
extract $a_{n}$ for any $n$, not necessarily positive integer. These
perturbative series then can be used to calculate the expansion of
$A_{n,m}$ to the desired order from the perturbative part of \eqref{eq:de1}.
By using these building blocks the all order solution for $q_{n,\kappa_{l}}$
, $W_{n,m}$ and $w_{n}$ can be built up.

The results for $q_{n,m}$, $W_{n,m}$ and $w_{n}$ are given in terms
of trans-series, which is understood as laterally Borel resummed.
This prescription does not follow from our derivation, although very
plausible from the contour shift, see also \citep{Marino:2021dzn}
for comments about this point. Thus we assume that the lateral Borel
resummation of the trans-series solution gives the TBA result. Since
the TBA result is free of ambiguities, we can calculate the various
alien derivatives of $A_{n,m}$ from the ambiguity cancellations\footnote{In this short note we assume that all Stokes constants, $S_{l}$ are
real. This is true for most of the models we analyse here. For complex
Stokes constants one has to take the appropriate real part of the
expressions, which complicates the discussion and we postpone the
detailed analysis for \citep{Bajnok:2022prepi}.}. These quantities differ only by the various source terms, thus we
expect that the $(\mathbf{I}-\mathbf{A})^{-1}$ operation guarantees
the ambiguity cancellation.

We analyze the behaviour of $q_{n,m}$ using resurgence theory and
alien derivatives following \citep{Dorigoni:2014hea,Aniceto:2018bis,Abbott:2020qnl}.
Assuming $m>0$ and $n>\kappa_{1}$, the leading singularity of the
Borel transform of $A_{n,-m}$ on the positive real line is at $\kappa_{1}$.
The corresponding ambiguity, which is encoded in the alien derivative
\footnote{The alien derivative is understood in the running coupling $v$ as:
$\dot{\Delta}_{n}=\nu^{n}\Delta_{n}$, where $[\dot{\Delta}_{n},\partial_{B}]=0$.
Thus it has an extra $v^{-\gamma}$ factor compared to the standard
definition.} $\Delta_{\kappa_{1}}A_{n,-m}$ has to be canceled by the leading
non-perturbative correction of order $\nu^{\kappa_{1}}$, i.e. by
$iS_{l}A_{n,-\kappa_{l}}A_{-\kappa_{l},-m}$ leading to $\Delta_{\kappa_{1}}A_{n,-m}=2iS_{1}A_{n,-\kappa_{1}}A_{-\kappa_{1},-m}$.
By moving iteratively further and subtracting the already known alien
derivatives one can show that 
\begin{equation}
\Delta_{\kappa_{l}}A_{n,m}=2iS_{l}A_{n,-\kappa_{l}}A_{-\kappa_{l},m}\quad.\label{eq:DeltaAnm}
\end{equation}
In a similar way one can also show that $\Delta_{n}A_{n,-m}=2iS_{0}A_{-n,-m}$
thus naturally extending the result for $l=0$.

We can then construct a multi-parameter trans-series for our basic
quantity as 
\begin{equation}
\hat{q}_{n,m}(\{\sigma\})=\sum_{\mathrm{paths}}sA_{n,\mathrm{path},-m}\sigma_{\mathrm{path}}\quad,
\end{equation}
where $\sigma_{\mathrm{path}}=\sigma_{l_{1}}\nu^{\kappa_{l_{1}}}\dots\sigma_{l_{N}}\nu^{\kappa_{l_{N}}}$.
One can even formally replace the $\sigma_{n}^{+}\nu^{n}A_{-n,-m}$
term with $iS_{0}\nu^{\kappa_{0}}\sigma_{0}A_{n,-m}$ and introduce
the trans-series parameter $\sigma_{0}$. As $\Delta_{n}^{2}q_{n,m}=0$,
the trans-series expression is linear in $\sigma_{0}$. By using \eqref{eq:qnmsol}
and \eqref{eq:DeltaAnm} one can show that the action of the pointed
alien derivative $\dot{\Delta}_{\kappa_{l}}=\nu^{\kappa_{l}}\Delta_{\kappa_{l}}$
on the trans-series is equivalent to $2iS_{l}$ times differentiation
wrt. $\sigma_{l}$: 
\begin{equation}
\dot{\Delta}_{\kappa_{l}}\hat{q}_{n,m}(\{\sigma\})=2iS_{l}\partial_{\sigma_{l}}\hat{q}_{n,m}(\{\sigma\})\quad.
\end{equation}
The Stokes automorphism which relates the two lateral Borel resummations
${\cal S}_{\pm}$ ( is the exponentiation of all the alien derivatives,
which then acts on the trans-series parameters as 
\begin{align}
\frak{S}\hat{q}_{n,m}(\{\sigma\}) & =e^{\sum_{l=0}\dot{\Delta}_{\kappa_{l}}}\hat{q}_{n,m}(\{\sigma\})=e^{\sum_{l=0}2iS_{l}\partial_{\sigma_{l}}}\hat{q}_{n,m}(\{\sigma\})\nonumber \\
 & =\hat{q}_{n,m}(\{\sigma_{l}\to\sigma_{l}+2iS_{l}\})\qquad.
\end{align}
The ambiguity free median resummation 
\begin{equation}
{\cal S}_{-}\frak{\frak{S}}^{\frac{1}{2}}\hat{q}_{n,m}(\{\sigma\})={\cal S}_{-}\hat{q}_{n,m}(\{\sigma_{k}\to\sigma_{k}+iS_{k}\})
\end{equation}
is then nothing but the TBA result, if we turn off every $\sigma$.
Note also that the inverse Stokes automorphism, which relates the
two lateral Borel resummations the opposite way is equivalent to $\sigma_{k}\to\sigma_{k}-iS_{k}$,
which corresponds to the alternative integrations in the contour deformations
as we anticipated before. Similarly one can show that $W_{n,m}={\cal S}_{-}(\frak{S}^{1/2}A_{m,n})$.
In particular, for the energy density we obtain $W_{1,1}={\cal S}_{-}({\cal \frak{S}}^{\frac{1}{2}}A_{1,1})$,
which is nothing but the median resummation of the perturbative series.

\section{Examples}

In this section we provide explicit examples for our generic solution
in various models.

There is a large class of integrable particle-models, where a magnetic
field can be coupled to one of the global charges and the energy density
can be investigated by the thermodynamic limit of the Bethe ansatz
equations. In these cases the kernel is related to the logarithmic
derivative of the scattering matrix and the Wiener-Hopf method leads
to a generic structure. We focus here on the bosonic models having
\begin{equation}
\sigma(i\kappa\pm0)=e^{\gamma\kappa\log\kappa+b\kappa}\frac{H(-\kappa)}{H(\kappa)}\left(\mp i\cos(\frac{\gamma\pi\kappa}{2})+\sin(\frac{\gamma\pi\kappa}{2})\right)\quad,\label{eq:sigma}
\end{equation}
where $\gamma,b$ are model-dependent constants, while $H(\kappa)$
is a model-dependent product of gamma-functions. We indicated the
signs of the residues depending on the two possible ways how we can
shift the cut away from the imaginary line. The poles on the imaginary
line are determined by the careful analysis of $H(\kappa)$, which
we go through model by model. Clearly, the running coupling can be
always introduced with $\gamma$ and $b$, together with another linear
term coming from the $H$-s, leading to the kernel
\begin{equation}
{\cal A}(x)=\cos(\frac{\gamma\pi vx}{2})e^{\gamma vx(\log x+q)+\sum_{k=1}^{\infty}z_{2k+1}(vx)^{2k+1}}\quad,
\end{equation}
where $z_{k}$ is proportional to $\zeta_{k}$ in a model-dependent
way and $q$ is parametrizing the various running couplings, which
we fix to the convenient value $q=\gamma_{E}+2\ln2$.

We start with the observable $A_{1,1}$, which can be calculated by
modifying Volin's method to keep track of the $\zeta$-s coming from
the kernel. The result is 
\begin{align}
2A_{1,1} & =1+\frac{v}{2}+\left(\frac{5\gamma}{4}+\frac{9}{8}\right)v^{2}+\left(\frac{10\gamma^{2}}{3}+\frac{53\gamma}{8}+\frac{57}{16}\right)v^{3}\\
 & +\frac{v^{4}}{384}\left(-36\gamma^{3}(21\zeta_{3}-94)+10924\gamma^{2}+13344\gamma+9(144z_{3}+625)\right)+O(v^{5}).\nonumber 
\end{align}
 We regard this as an input to our analysis and show how all the perturbative
parts can be determined from this. We can calculate this series analytically
up to 50 orders while numerically up to few hundred orders for generic
$N$ and up to 2000 terms for $N=4$ with very high precision \citep{Abbott:2020qnl,Abbott:2020mba,Bajnok:2021zjm}.
For demonstration, we merely included here the first few terms, and
keep doing the same from now on.

By using the differential equation $2A_{1,1}+\dot{A}_{1,1}=a_{1}^{2}$,
i.e. the perturbative part of \eqref{eq:de1}, one can obtain
\begin{align}
a_{1} & =1+\frac{v}{4}+\left(\frac{5\gamma}{8}+\frac{9}{32}\right)v^{2}+\left(\frac{5\gamma^{2}}{3}+\frac{53\gamma}{32}+\frac{75}{128}\right)v^{3}\\
 & +\frac{v^{4}}{6144}\left(9(1152z_{3}+1225)-288\gamma^{3}(21\zeta_{3}-94)+43696\gamma^{2}+35160\gamma\right)+O(v^{5}).\nonumber 
\end{align}
Then using \eqref{eq:de2} for $n=1$ we obtain 
\begin{equation}
f=-v^{2}-6\gamma v^{3}-26\gamma^{2}v^{4}+v^{5}\left(\frac{1}{4}\gamma^{3}(63\zeta_{3}-386)-27z_{3}\right)+O\left(v^{6}\right)\quad.
\end{equation}
By solving \eqref{eq:de2} for other $n$-s we can get 
\begin{align}
a_{n} & =1+\frac{v}{4n}+\frac{v^{2}(20\gamma n+9)}{32n^{2}}+\frac{v^{3}\left(640\gamma^{2}n^{2}+636\gamma n+225\right)}{384n^{3}}\\
 & \quad+\frac{v^{4}\left(288n^{3}\left(\gamma^{3}(94-21\zeta_{3})+36z_{3}\right)+43696\gamma^{2}n^{2}+35160\gamma n+11025\right)}{6144n^{4}}+O\left(v^{5}\right)\quad.\nonumber 
\end{align}
Actually $a_{n}$ can be obtained directly from $a_{1}$ by the $v\to\frac{v}{n}$,
$\gamma\to\gamma n$, $z_{2k+1}\to n^{2k+1}z_{2k+1}$ replacements.
The exceptional $\chi_{0}$ is 
\begin{equation}
\chi_{0}=\frac{1}{\sqrt{v}}\left(1-\frac{\gamma v}{2}-\frac{5\gamma^{2}v^{2}}{8}+\frac{1}{16}v^{3}\left(\gamma^{3}(7\zeta_{3}-15)-12z_{3}\right)+O(v^{4})\right)\quad.
\end{equation}

Finally, by solving \eqref{eq:de1} we obtain the basic building blocks
\begin{align}
A_{n,m} & =\frac{1}{m+n}+\frac{v}{4mn}+\frac{v^{2}(20\gamma mn+9m+9n)}{32m^{2}n^{2}}\\
 & \quad+\frac{v^{3}\left(m^{2}\left(640\gamma^{2}n^{2}+636\gamma n+225\right)+6mn(106\gamma n+39)+225n^{2}\right)}{384m^{3}n^{3}}+O\left(v^{4}\right)\quad.\nonumber 
\end{align}
In order to get the generic solutions in terms of these $A$-s as
(\ref{eq:qnmsol},\ref{eq:Wnm},\ref{eq:wn}) we need the locations
$\kappa_{l}$ and Stokes constants $S_{l}$, which should be found
model by model.

\subsection{$O(N)$ models}

The $O(N)$ non-linear sigma models in a magnetic field coupled to
one of the $O(N)$ charges \citep{Hasenfratz:1990ab,Hasenfratz:1990zz}
can be analyzed by the thermodynamic limit of the Bethe Ansatz equation,
which takes the form of the integral equation (1) with the kernel
related to the S-matrix \citep{Zamolodchikov:1977nu}
\begin{equation}
S(\theta)=-\frac{\Gamma(\frac{1}{2}-\frac{i\theta}{2\pi})\Gamma(\Delta-\frac{i\theta}{2\pi})\Gamma(1+\frac{i\theta}{2\pi})\Gamma(\Delta+\frac{1}{2}+\frac{i\theta}{2\pi})}{\Gamma(\frac{1}{2}+\frac{i\theta}{2\pi})\Gamma(\Delta+\frac{i\theta}{2\pi})\Gamma(1-\frac{i\theta}{2\pi})\Gamma(\Delta+\frac{1}{2}-\frac{i\theta}{2\pi})}
\end{equation}
 as $K(\theta)=\frac{1}{2\pi i}\partial_{\theta}\log S(\theta)$,
where $\Delta=\frac{1}{N-2}$. For the $\cosh n\theta$ source term
${\cal O}_{n,m}$ describes the expectation value of the spin $m$
conserved charge with the Hamiltonian being the spin $n$ charge.
The energy density analyzed in the literature corresponds to ${\cal O}_{1,1}$.
The Wiener-Hopf decomposition gives \eqref{eq:sigma} with 
\begin{equation}
\gamma=2\Delta-1\quad;\quad H(\kappa)=\frac{\Gamma(1+\Delta\kappa)}{\Gamma(\frac{1}{2}+\frac{\kappa}{2})}\quad;\quad b=-2\Delta(1-\ln\Delta)+(1+\ln2)
\end{equation}
 and the kernel is described by 
\begin{equation}
z_{2k+1}=2\frac{\zeta_{2k+1}}{2k+1}(\Delta^{2k+1}-1+2^{-2k-1})
\end{equation}
in terms of the running coupling \ref{eq:runningcoupling} defined
with the specific choice $L=b-4\Delta\ln2$ for the arbitrary constant.
The zeros of $\sigma(i\kappa)$ are located $N$-independently at
the positions $\kappa=2l-1$, while its poles are at $\kappa=l(N-2)$,
where $l\in\mathbb{N}$. This implies that $\kappa_{l}=l\kappa_{1}$
with $\kappa_{1}=N-2$ for $N$ even and $\kappa_{1}=2N-4$ for $N$
odd \citep{Marino:2021dzn,Bajnok:2022rtu}.

\subsubsection*{$O(4)$ model}

\begin{figure}
\begin{centering}
\includegraphics[width=12cm]{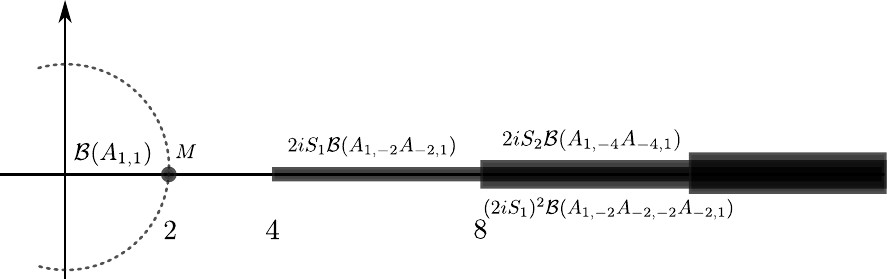}
\par\end{centering}
\caption{Analytic structure of the Borel transform of $A_{1,1}$, ${\cal B}(A_{1,1})$,
with its convergence radius indicated. The first singularity is a
pole at $2$ with residue $M$. The second singularity is a logarithmic
cut starting at $4$ multiplied with a function whose perturbative
expansion is $2iS_{1}A_{1,-2}^{2}$. This function also has a logarithmic
cut starting at $4$ with the function $(2iS_{1})^{2}{\cal B}(A_{1,-2}^{2}A_{-2,-2})$
multiplying it. This cut in ${\cal B}(A_{1,1})$ shows up at $8$,
which coincides with another logarithmic cut with function $2iS_{2}{\cal B}(A_{1,-4}^{4})$.
These two cuts can be disentangled by the different reality properties.}

\label{figure}
\end{figure}

Let us start with the $O(4)$ model, which is the simplest. In this
case the running coupling is $v=\frac{1}{2B}-2\ln2$ and $\Delta=\frac{1}{2}$.
The poles and the zeros do not interact and $\kappa_{l}=2l$; $l=1,2,\dots$,
with residues 
\begin{equation}
S_{l}=\frac{((2l-1)!!)^{2}}{2^{2l-1}l!(l-1)!}\quad.
\end{equation}
Observe also that $\sigma(in)=0$ for $n$ odd, i.e. $S_{0}=0$, so
in these cases $\kappa_{0}=n$ is not singular and we do not have
the $l=0$ term in the sums. With these building blocks the trans-series
for the observable $w_{1}$ takes the form 
\begin{equation}
w_{1}=a_{1}+\sum_{l_{1},l_{2},\dots}e^{-4(l_{1}+l_{2}+\dots)B}(iS_{l_{1}})(iS_{l_{2}})\dots A_{1,-2l_{1}}A_{-2l_{1},-2l_{2}}\dots a_{-l_{k}}\quad.
\end{equation}
 It is free of ambiguities due to the relation 
\begin{equation}
\Delta_{2l}A_{\alpha,\beta}=2iS_{l}A_{\alpha,-2l}A_{-2l,\beta}\quad.
\end{equation}
Finally $W_{1,1}$ can be obtained from eq. \eqref{eq:de1} as
\begin{equation}
W_{1,1}=A_{1,1}+Me^{-2B}+\sum_{l_{1},l_{2},\dots}e^{-4(l_{1}+l_{2}+\dots)B}iS_{l_{1}}iS_{l_{2}}\dots A_{1,-2l_{1}}A_{-2l_{1},-2l_{2}}\dots A_{-l_{k},1}\quad,
\end{equation}
where $M$ is an integration constant, which comes from the zero mode
of $2+\partial_{B}$. By taking the $n\to1$ and $m\to1$ limit in
\eqref{eq:Wnm} it can be calculated explicitly to be $M=-2i$. The
first few terms take the form
\begin{equation}
W_{1,1}=A_{1,1}+\frac{M}{2}e^{-2B}+ie^{-4B}S_{1}A_{1,-2}^{2}+e^{-8B}((iS_{1})^{2}A_{1,-2}^{2}A_{-2,-2}+iS_{2}A_{1,-4}^{2})+\dots\quad.
\end{equation}
By explicitly investigating the analytic structure of $A_{1,1}$ on
the Borel plane (displayed on Figure \ref{figure}) we confirmed the
perturbative expansion of all these terms up to high orders. We also
verified numerically that the median resummation reproduced the TBA
result.

These results have a direct extension for $N>4$. The only difference
is that $\kappa_{1}$ is $N$-dependent, otherwise the poles form
the lattice $\kappa_{l}=l\kappa_{1}$ and the generic solutions in
terms of the $A$-s in eqs. (\ref{eq:qnmsol},\ref{eq:Wnm},\ref{eq:wn})
applies. The integration constant for $W_{1,1}$ is $M=-2e(\frac{\Delta}{e})^{2\Delta}\frac{\Gamma(1-\Delta)}{\Gamma(1+\Delta)}e^{i\pi\Delta}$.

\subsubsection*{$O(3)$ model}

The $O(3)$ model is the most complicated among the $O(N)$ models.
This is due to the fact that $\sigma(i)\neq0$, and we have to carry
the $l=0$ term in the sums for $W_{1,1}$ and $w_{1}$. The poles
of $\sigma(i\kappa)$ are again located at $\kappa_{l}=2l$. The building
blocks $A_{n,m}$ can be used here with $\Delta=1$. By focusing on
$w_{1}$ the main difference compared to the $O(4)$ model is that
additionally to the $O(4)$ like sums, we also have others starting
at $\nu$: 
\begin{align}
w_{1} & =a_{1}+\sum_{l_{1},l_{2},\dots}\nu^{2(l_{1}+l_{2}+\dots)}iS_{l_{1}}iS_{l_{2}}\dots A_{1,-2l_{1}}A_{-2l_{1},-2l_{2}}\dots a_{-l_{k}}\\
 & \quad iS_{0}\nu\biggl(1+\sum_{l_{1},l_{2},\dots}\nu^{2(l_{1}+l_{2}+\dots)}iS_{l_{1}}iS_{l_{2}}\dots A_{-1,-2l_{1}}A_{-2l_{1},-2l_{2}}\dots a_{-l_{k}}\biggr)\quad.\nonumber 
\end{align}
What is interesting is that the two parts are not related by any resurgence
relations, i.e. the $\Delta_{1}$ alien derivative of the first line
is not related to the second line. This can be also seen by noting
that $S_{0}$ is imaginary and the real leading term cannot be related
to the purely imaginary alien derivative of a real series. The even
alien derivatives satisfy the relations as before $\Delta_{2l}A_{n,m}=2iS_{l}A_{n,-2l}A_{-2l,m}$.
The observable $W_{1,1}$ can be obtained by integrating the differential
equation \eqref{eq:de1}. Again the constant term should be fixed
from taking the $n\to1$ and $m\to1$ limit.

\subsection{Principal chiral models}

The $SU(N)$ principal chiral model can be described by 
\begin{equation}
\gamma=0\quad;\quad H(\kappa)=\frac{\Gamma(1+(1-\Delta)\kappa)\Gamma(1+\Delta\kappa)}{\Gamma(1+\kappa)}\quad;\quad b=2\Delta\ln\Delta+2(1-\Delta)\ln(1-\Delta),
\end{equation}
where $\Delta=1/N$. In order to use the generic forms we need the
replacements
\begin{equation}
z_{2k+1}=2\frac{\zeta_{2k+1}}{2k+1}\left(-1+\Delta^{2k+1}+(1-\Delta)^{2k+1}\right)
\end{equation}
and the running coupling is defined with $L=b.$ The poles of $\sigma(i\kappa)$
again form a lattice $\kappa_{l}=l\kappa_{1}$ with $\kappa_{1}=\frac{N}{N-1}$.
This model is very similar to the $O(4)$ model, which is the $SU(2)$
case here.

\subsection{Supersymmetric $O(N)$ models}

In this model we have 
\begin{align}
\gamma & =-1\quad;\quad H(\kappa)=\frac{\Gamma(\frac{1}{2}+\frac{(1-2\Delta)\kappa}{2})\Gamma(1+\Delta\kappa)}{\Gamma(\frac{1}{2}+\frac{\kappa}{2})^{2}}\\
b & =(1+2\Delta)\ln2+2\Delta\ln\Delta+(1-2\Delta)\ln(1-2\Delta)+1,\nonumber 
\end{align}
where $\Delta=1/(N-2)$ with $N\geq5$ and 
\begin{equation}
z_{2k+1}=2\frac{\zeta_{2k+1}}{2k+1}\left(\Delta^{2k+1}-2+2^{-2k}+(1-2\Delta)^{2k+1}(1-2^{-2k-1})\right)\quad.
\end{equation}
The running coupling is defined with $L=b-4\Delta\ln2$. The first
case is $N=5$ for which $\kappa_{1}=6$. For $N>5$ we have to distinguish
between the even and odd cases just as we did for the $O(N)$ models.
We actually have the same set as for the $O(N)$ models and additionally
$\mu_{l}=\frac{N-2}{N-4}(2l-1)$, although for odd $N$s some of the
residues are zero. This is a very complicated pattern and, additionally,
the Stokes constants are complex. We are planning to investigate these
cases in detail in our forthcoming publication \citep{Bajnok:2022prepi}.

\section{Conclusion}

In this paper we developed a method to solve completely the integral
equations \eqref{eq:TBA} in terms of a trans-series. By taking the
perturbative energy density $A_{1,1}$ as an input we determined a
set of observables $A_{n,m}$ which constitute a complete basis in
the trans-series solution. We used these building blocks to construct
the full trans-series for various other observables including the
generalized energy densities and the boundary values of the Bethe
Ansatz densities. We also revealed the analytical structure of all
$A_{n,m}$-s on the Borel plane by determining their (positive) alien
derivatives. The singularities on the positive real lines are interplayed
such a way that the TBA result agrees with the median resummation.
We checked some of our calculations with the explicit examples of
the bosonic integrable models, in particular of the $O(N)$ non-linear
sigma models and this provides further support to our basic assumption
of lateral Borel resummation. To our knowledge these are the first
explicitly solved asymptotically free quantum field theories.

In the statistical physical applications the systems are not relativistically
but Galiean-invariant. This implies that the source terms and the
moments has to be changed from $\cosh n\theta$ to $\theta^{j}$,
see \citep{PhysRevA.106.062216,PhysRevLett.130.020401} for details
in constructing the analogue of our basis and for describing observables
at the perturbative level. Our formulas can be related to those by
differentiating wrt. $n$ and putting $n$ to zero. However, here
we go beyond the perturbative level, and construct the full non-perturbative
trans-series. We think that by making the appropriate differentiation
the non-perturbative parts of the non-relativistic moments can also
be extracted based on our formulae.

We thus hope that our generic solution of the TBA equation will be
fruitful and pave the way of further applications both in statistical
and particle physics.

\subsection*{Acknowledgments}

ZB thanks Ines Aniceto the useful discussions. ZB and JB would like
to thank the Isaac Newton Institute for Mathematical Sciences, Cambridge,
for support and hospitality during the programme \emph{Applicable
resurgent asymptotics: towards a universal theory} where work on this
paper was undertaken. This work was supported by EPSRC grant no EP/R014604/1
and NKFIH research Grant K134946.

\bibliographystyle{elsarticle-num}
\bibliography{paper}

\end{document}